\documentclass[aps,prl,reprint,superscriptaddress]{revtex4-2}

\pdfoutput=1 %% Uncomment for arxiv Submission
\usepackage{amsmath}
\usepackage{amssymb}
\usepackage{graphicx}% Include figure files
\usepackage{bm}% bold math
\usepackage{braket}
\usepackage{hyperref}
\usepackage[usenames]{xcolor}
\usepackage[capitalize]{cleveref}
\usepackage{siunitx}
\usepackage{physics}

\begin{document}

\title{Low-frequency Quantum Oscillations from Interactions in Layered Metals}

\author{Andrew A. Allocca} \email{aa2182@cam.ac.uk}
\affiliation{T.C.M. Group, Cavendish Laboratory, University of Cambridge, JJ Thomson Avenue, Cambridge, CB3 0HE, U.K.}
\author{Nigel R. Cooper}
\affiliation{T.C.M. Group, Cavendish Laboratory, University of Cambridge, JJ Thomson Avenue, Cambridge, CB3 0HE, U.K.}
\affiliation{Department of Physics and Astronomy, University of Florence, Via G. Sansone 1, 50019 Sesto Fiorentino, Italy}

\date{\today}

\begin{abstract}
Metals composed of weakly-coupled, stacked layers possess a Fermi surface that slightly varies in size along the stacking direction.
This appears in  de Haas-van Alphen (dHvA) oscillations of the magnetisation with magnetic field as two close frequencies, corresponding to the two extremal Fermi surface cross-sectional areas.     
We show that, for layered materials of sufficiently high mobility, Coulomb interactions can have a dramatic effect on the form of the dHvA oscillations: there is also generically an oscillation at the small difference of the two large frequencies. 
We determine the size and form of this effect, and show that it probes the short-range part of the Coulomb interactions within the layered material. 
We argue that this interaction effect may explain recent experimental observations of anomalous low-frequency dHvA oscillations in the ultrapure delafossites.
\end{abstract}

\maketitle

\emph{Introduction} -- Modern materials science has produced a wide variety of new sorts of solid state systems, such as those characterized by non-trivial topology or strong interactions. The novel properties of these materials have  led to reexaminations of previously well-understood phenomena.
Quantum oscillations (QOs) of the magnetization as a function of (inverse) magnetic field~\cite{deHaas1930}, long interpreted in terms of the geometry of closed Fermi surfaces, and the standard Lifshitz-Kosevich (LK) theory~\cite{Lifshitz1956}, have been necessarily reexplored following the discovery of Weyl and nodal semimetals~\cite{Potter2014, Arnold2016, Alexandradinata2017, Nair2020,Devakul2021}, quasiperiodic systems~\cite{Zhang2015, Spurrier2019}, and the observation of QOs in insulators~\cite{Tan2015, Knolle2015, Baskaran2015, Onur2016, Zhang2016, Knolle2017, Sodemann2018, Hartstein2018}, systems all featuring unusual or absent Fermi surfaces. 

Following in this vein, recent de Haas-van Alphen (dHvA) studies on delafossites, a class of layered materials featuring strong interactions and high in-plane mobility, also display anomalous behavior---large, low-frequency QOs at the difference of the two natural high frequencies related to extremal Fermi surface areas~\cite{Hicks2012, Arnold2020}. These observations are not readily explained by existing theory based on magnetic interactions~\cite{Shoenberg1984}, and require a re-examination of the theory of the dHvA effect in these new materials settings.

In this Letter, we show that Coulomb interactions can lead to difference-frequency oscillations of the magnetization, of a size that can account for these experimental observations. 
The mechanism that we identify is overlooked in long-standing theories of interaction effects on the dHvA effect~\cite{Luttinger1961,Shoenberg1984,Wasserman1996}, as it requires the retention of terms that are smaller than those needed to account  for the high-frequency oscillations. However, in high-mobility systems such as the delafossites, the dominant dHvA oscillation at the difference frequency in both bilayer and multi-layer systems can arise from the Coulomb interaction mechanism we identify.
In these circumstances our theory shows that the measurement of a difference-frequency oscillation is a signature of strong non-local interactions. Our results indicate that, in general, new considerations are warranted when analyzing dHvA oscillations in high-mobility metals.

To trace the origin of a difference frequency in QOs, first consider a non-interacting Fermi gas whose Fermi surface has two extremal orbits of similar areas. 
As the magnetic field varies, the Landau quantized orbits sweep through these two extremal areas, causing thermodynamic quantities to  oscillate at the two frequencies $f_\pm$. 
Within a model of non-interacting electrons at fixed chemical potential these oscillations are independent. 
However, any nonlinearities that couple these two oscillations at $f_\pm$, can lead to a difference frequency $\delta f = |f_+-f_-|$ (as well as side-bands). 
One source of nonlinearity is the magnetic interaction, by which the magnetic field $B = \mu_0 (H+M)$ acquires an oscillatory component through the oscillating magnetization $M(H)$. 
This leads to a difference frequency component of size related to the geometry-dependent demagnetization field~\cite{Shoenberg1984}. However, this effect appears insufficient to explain the difference-frequency oscillations of magnetization seen in recent experiments~\cite{Hicks2012, Arnold2020}. 

As we will show, a much larger difference frequency oscillation can arise from Coulomb interactions. 
One consequence of the Coulomb interaction is that its long-range component forces overall charge neutrality in bulk 3D materials, requiring one to work at fixed electron number density $n$ rather than fixed chemical potential.
This introduces a non-linearity, mediated by oscillations in the chemical potential~\cite{Champel2001, Grigoriev2001}, that can readily produce a large difference frequency oscillation in the conductivity (Shubnikov-de Hass effect)~\cite{Grigoriev2003}. 
While we find that this can also produce a difference frequency oscillation in the magnetization (i.e. in the dHvA effect), this effect by itself is very small. In fact, as we will show, the dominant effect arises from a full consideration of the Coulomb interactions that also takes account of the (screened) short-range interactions that couple local fluctuations of the charge density.
The combination of both short- and long-range components of the Coulomb interaction produces a much larger difference frequency oscillation than either can independently.

\emph{Bilayer (2D)} -- To illustrate the essential physics underlying this mechanism, we first consider a bilayer model. 
(We neglect possible strongly correlated phases that can arise from strong interactions in 2D Landau quantized systems.) 
We consider a pair of identical, parallel 2D gases of spinless electrons, with parabolic in-plane dispersion of effective mass $m^\ast$. 
Interlayer hopping, of amplitude $\Delta/2$, splits the energy eigenstates into symmetric and antisymmetric subbands, and a perpendicular magnetic field $B$ reorganizes the in-plane states into Landau levels. 
The spectrum is  then $\epsilon_{l,s} = \hbar\omega_c(l+1/2) - s \Delta/2$, with $\omega_c = eB/m^\ast$ the cyclotron frequency, $l = 0,1,2,\dots$ indexing Landau levels, and $s=\pm$ marking even and odd subbands. 
For a chemical potential $\mu$, the densities of the two subbands at $B=0$ are $n_\pm(0) = (\mu\pm \Delta/2)m^*/(2\pi\hbar^2) $ and there are two distinct Fermi surfaces.
At non-zero $B$ these give rise to QOs in the magnetization at frequencies  $f_\pm = en_\pm(0)/h = m^\ast(\mu \pm \Delta/2)/e\hbar$. 
Within LK theory these are the only two frequencies in the magnetization oscillations. 

At fixed chemical potential $\mu$, the occupations of the two subbands, $n_\pm(B)$, both  oscillate with $B$. 
For $\hbar\omega_c \ll \Delta \ll \mu$, so that many Landau levels are occupied and $n_\pm$ have similar average values, these oscillations are on the scale of $n_\Phi = B/\Phi_0$, the degeneracy of each Landau level per unit area, with $\Phi_0 = h/e$ the magnetic flux quantum. 
Keeping only the first harmonic, we write 
\begin{gather}
\label{eq:dens}
n_\pm(B) \sim n_\pm(0)  - \eta {n_\Phi}\sin\left(2\pi \frac{\mu \pm \Delta/2}{\hbar\omega_c}\right)
\end{gather}
where $\eta \lesssim 1$ is introduced phenomenologically  to account for any disorder and temperature effects that suppress oscillations. 

The oscillations in density \cref{eq:dens} lead to oscillations in the Coulomb interactions that can also have components at the difference frequency. Taking the ionic background charge density to be $n_I$ in each well, and computing the expectation value of the electron-electron interactions in the non-interacting ground state leads to the interaction energy
\begin{equation}
\label{eq:int}
    E_{\text{int}} \sim V [\bar{n}(B)-n_I]^2 - V [\delta n(B)/2]^2,
\end{equation}
with $\bar{n} \equiv (n_++n_-)/2$ and $\delta n \equiv n_+ - n_-$.  The first term in (\ref{eq:int}) is the Hartree energy, related to the overall electrostatic energy of fluctuations of the total charge density.  
The second term is the Fock term, describing the suppression of short-range repulsion due to exchange~\footnote{The numerical values of the prefactors of these  two terms can  vary, depending on the spatial form of the inter-electron interactions. 
The form in (\ref{eq:int}) arises for the model interactions chosen in (\ref{eq:sint}), of short-range inter-layer interactions.}.
Since  the subband densities $n_\pm$ oscillate at the frequencies $f_\pm$, the cross term $n_+ n_-$ produces a  small component that oscillates at the difference frequency $\delta f = \abs{f_+-f_-}$, with amplitude  of order $V \eta^2 n_\Phi^2$.

To account for the effects of the long-range part of the Coulomb interaction, we enforce  charge neutrality, requiring   $\bar{n}(B) \equiv n_I$.
The chemical potential $\mu$  becomes an oscillating function of $B$, which to first order in  $\eta$ is
\begin{equation} \label{eq:mu2D}
     \mu(B)  \approx \mu(0) + \frac{\eta \hbar\omega_c}{2} \sum_{s=\pm}\sin\left(2\pi\frac{\mu(0) + s\Delta/2}{\hbar\omega_c}\right).
%    \Rightarrow \theta_1(B) = -\frac{1}{2} \sum_{s=\pm}\sin\left(2\pi\frac{f_s}{B}\right).
\end{equation}
With this oscillating $\mu(B)$ inserted in \cref{eq:dens} one finds that the density $n_\pm(B)$ oscillates not just at $f_\pm$ but also at the difference frequency $\delta f$ (and at other sidebands).
The restriction to fixed $n$ has a large effect on the interaction energy. 
The Hartree term vanishes exactly, leaving just the Fock exchange energy, $-V[\delta n/2]^2$, which contains an oscillatory term  of size
\begin{equation} \label{eq:Eintn}
  \sim V\eta n_\Phi \delta n(0) \sum_{s=\pm}s\sin\left(2\pi\frac{\mu(B)+s \Delta/2}{\hbar\omega_c}\right)  + \ldots
\end{equation}
This  gives rise to oscillations at the difference frequency $\delta f$, with amplitude of order  $V \eta^2  n_\Phi \delta n (0)$. 
This is larger than the $\delta f$ oscillation we identified at fixed chemical potential by a factor of $\delta n (0)/n_\Phi = \Delta / (\hbar\omega_c)$, which can be large at low fields.

\emph{Multi-layer (3D)} -- We now turn to a model for a multi-layer metal. As above, each layer is described by a parabolic dispersion with effective mass $m^\ast$, electrons may hop between adjacent layers with an amplitude that we now denote $t_\perp$, and a perpendicular field $B$ breaks in-plane states into Landau levels.
At nonzero temperature $T$ the system is described by the action
\begin{equation}
     S_0 = \sum_{\epsilon_n,l,k_y}\int_{k_z} \!\! \bar{\psi}_{l,k_y}(k_z,\epsilon_n)\left(-i\epsilon_n +\xi_l(k_z)\right) \psi_{l,k_y}(k_z,\epsilon_n),
\end{equation}
where $\psi$, $\bar{\psi}$ are the electron field operators in the energy eigenbasis, $\epsilon_n = (2n+1)\pi k_B T$ is the Matsubara frequency, and $\xi_l(k_z) = \hbar\omega_c(l+1/2) - 2t_\perp \cos(k_z a_\perp) - \mu$ is the single-particle energy measured from the Fermi level.
The discrete eigenstate index $s$ of the bilayer is replaced by the continuous quasi-momentum $k_z \in (-\pi/a_\perp,\pi/a_\perp]$ describing dispersion along the $c$-axis, and we use the notation $\int_{k_z} = \int_{-\pi/a_\perp}^{\pi/a_\perp}dk_z/2\pi$, where $a_\perp$ is the interlayer spacing.
With our choice of Landau gauge, $k_y$ indexes the degenerate states in each Landau level with total number $\sum_{k_y} =  A n_\Phi$, with $A$ the sample area.
The frequencies determined by the two extremal cross-sectional areas of the Fermi surface along $k_z$ are $f_\pm = m^\ast(\mu \pm 2t_\perp)/e\hbar$~\footnote{We extract numerical values for $\mu$ and $t_\perp$ by fitting the frequencies reported in Ref.~\cite{Arnold2020} to this form.}. %, so the difference frequency $\delta f = \abs{f_+-f_-} = 4m^\ast t_\perp/e\hbar$ is twice as large here as for the bilayer.
As before we consider spinless electrons, as this already shows the new effect.

For simplicity, in considering the interlayer interaction between local charge density fluctuations we approximate the (screened) Coulomb potential as only acting between pairs of nearest points on neighboring layers, $V_{i,i'}(\mathbf{r-r'}) = \delta_{i',i+1} e^2/4\pi\varepsilon \abs{\mathbf{r-r'}+a_\perp\mathbf{\hat z}} \approx V \lambda^2 \delta_{i',i+1} \delta(\mathbf{r-r'})$, where $\varepsilon$ is the permittivity and $\lambda^2$ is the area of the ``patch'' on each layer that participates in the interaction. %area ``occupied'' by each electron in-plane.
The interaction term in the action is
\begin{equation}
\label{eq:sint}
    S_\text{int} \approx V\lambda^2\!\! \int_0^\beta \!\!\!d\tau\!\! \int\!\! d\mathbf{r} \sum_i \bar{c}_i(x) \bar{c}_{i+1}(x) c_{i+1}(x) c_i(x),
\end{equation}
where %$\tau$ is imaginary time, $\beta = 1/k_BT$ is the inverse temperature, and 
$x = (\tau,\mathbf{r})$.
We also acquire a term $S_I = \beta AL_z V \lambda^2 n_I(n_I/a_\perp - 2n)$, where $L_z$ is the extent of the system in $z$, which depends only on the total electron density $n$ and accounts for the interaction between electrons and lattice ions.
Note that here $n$ represents a number per unit 3D volume, whereas $n_I$ is a number per unit 2D area in-plane.

To analyze the interaction effects we proceed as follows.
The interlayer interaction can be included to first order with the Hartree-Fock self-energy $\Sigma$, inserted into the full electron Green's function, $G = \left(G_0^{-1} - \Sigma\right)^{-1}$, with $G_0 = (i\epsilon_n-\xi)^{-1}$ the free electron Green's function.
It is important for our analysis to keep both the constant, zero-field part $\Sigma_0$ and oscillatory part $\widetilde{\Sigma}(B)$ of the self-energy;
previous studies of the dHvA effect including interactions have used the general relation $\widetilde{\Sigma} \ll \Sigma_0$ to discard $\widetilde{\Sigma}$ entirely~\cite{Luttinger1961,Shoenberg1984,Wasserman1996}, but we find that this term produces the leading contribution to difference frequency oscillations in this sort of layered system.

From the Green's function $G$ we calculate the grand potential $\Omega$ via standard field theoretic methods, expanding up to first order in the interaction constant $V$. % and putting $\Omega \approx \Omega^{(0)} + \Omega^{(1)}$, where $\Omega^{(0)}$ is the contribution for the non-interacting system and $\Omega^{(1)}$ contains all terms first order in $V$.
We then obtain the free energy $F$ as the Legendre transform of the grand potential, $F(B,n) = \Omega(B,\mu(B,n)) + n\mu(B,n)$, where $n$ is the fixed electron density, and $\mu(B,n)$ is the oscillatory chemical potential needed to fix the electron density.
To linear order in $V$, it is sufficient to  fix the density using just the non-interacting part of the theory, i.e.\ defining $\mu(B,n)$  through $n = -\partial\Omega/\partial\mu\vert_{V=0}$; corrections to $\mu$ that depend on $V$ only lead to terms in $F$ that are of order $V^2$ and higher~\cite{Supplement}. 
The result of this procedure--the free energy $F(B,n)$--is the relevant thermodynamic potential for a system with fixed electron density, including all first order interlayer interaction effects. 

We calculate the Hartree-Fock self-energy in the approximation that the interaction is independent of Landau-level index,
\begin{gather}
    \Sigma(k_z,B) = V\lambda^2 a_\perp \left(n(B) - \chi(B) \cos(k_z a_\perp)\right)\\
    n(B) = \!\int_{k_z}\!\! n(k_z,B), \,\,
    \chi(B) = \!\int_{k_z}\!\! \cos(k_z a_\perp) n(k_z,B).
\end{gather}
This gives the exact form of $\Omega$ up to first order in $V$.
Here we define the 3D number density of occupied states at momentum $k_z$, $n(k_z,B) = n_\Phi \int\! d\epsilon \sum_l n_F(\epsilon) \mathcal{A}(\epsilon-\xi_l(k_z))$, where $\mathcal{A}$ is the spectral density.
With the 3D electron density fixed to $n(B) = n = n_I/a_\perp$ we determine the oscillatory chemical potential $\mu(B,n)$~\cite{Supplement}, then obtain the free energy $F_\text{3D} = F_\text{3D}^{(0)} + F_\text{3D}^{(1)}$ with
\begin{gather} 
    F_\text{3D}^{(0)} = n_\Phi\int_{k_z}\!\int\!\! d\epsilon \,\epsilon\, n_F(\epsilon) \sum_{l}\mathcal{A}(\epsilon-\xi_l(k_z)) + n\,\mu(B,n) \label{eq:F3D0} \\
    F_\text{3D}^{(1)} = -V\lambda^2a_\perp \chi(B,\mu(B,n))^2. \label{eq:F3D1}
\end{gather}
As in the 2D case, %kinetic and interacting contributions to the energy cleanly separate, and 
the interaction part $F_\text{3D}^{(1)}$ is given solely by the Fock energy.

We analytically evaluate the oscillatory part of the free energy assuming the hierarchy of energy scales $\hbar\omega_c \ll 2t_\perp \ll \mu$. Equivalently this sets a hierarchy of 2D densities $n_\Phi \ll n_\perp \ll n_I$, where we define $n_\perp \equiv t_\perp m^*/\pi \hbar^2$.
The dominant contributions to oscillations at $f_\pm$ and $\delta f$ are found to be~\cite{Supplement} 
\begin{multline} \label{eq:F3D0osc}
    \widetilde{F}_\text{3D}^{(0)} = \frac{\hbar\omega_c n_\Phi^2}{8\pi^4 a_\perp n_\perp} R_{D,1}^2R_{T,1}^2 \sin\left(\frac{2\pi\,\delta f}{B}\right)\\ 
    -\frac{\hbar\omega_c n_\Phi}{4\pi^3a_\perp}\sqrt{\frac{n_\Phi}{n_\perp}}R_{D,1}R_{T,1}\sum_{\alpha=\pm}\cos\left(\frac{2\pi f_\alpha}{B}-\frac{\alpha\pi}{4}\right) 
\end{multline}
from the kinetic part, and
\begin{multline} \label{eq:F3D1osc}
    \widetilde{F}_\text{3D}^{(1)} = V\lambda^2 \frac{n_\Phi^2}{2\pi^3a_\perp}R_{D,1}^2 R_{T,1}^2\cos\left(\frac{2\pi\,\delta f}{B}\right) \\
    + V\lambda^2 \frac{n_\Phi n_\perp}{2\pi^2a_\perp}\sqrt{\frac{n_\Phi}{n_\perp}}R_{D,1}R_{T,1}\sum_{\alpha=\pm} \alpha \sin\left(\frac{2\pi f_\alpha}{B}-\frac{\alpha\pi}{4}\right)
\end{multline}
from the interacting part, where
\begin{equation}\label{eq:Rfactors}
    R_{D,p} = \exp\left[-\frac{\pi p}{\omega_c \tau_\text{qp}}\right], \, R_{T,p} = \frac{2\pi^2 p\, \frac{k_B T}{\hbar\omega_c}}{\sinh\left(2\pi^2 p\, \frac{k_B T}{\hbar\omega_c}\right)},
\end{equation}
are the Dingle factor and LK temperature factor, accounting for finite quasiparticle lifetime $\tau_\text{qp}$ and nonzero temperature $T$ respectively.
Note that in both Eqs.~\eqref{eq:F3D0osc} and \eqref{eq:F3D1osc} the $\delta f$ terms are found to depend on the square of these factors, while the $f_\pm$ terms only depend on a single power of each. 

\begin{figure}[t]
    \centering
    \includegraphics[width=\columnwidth]{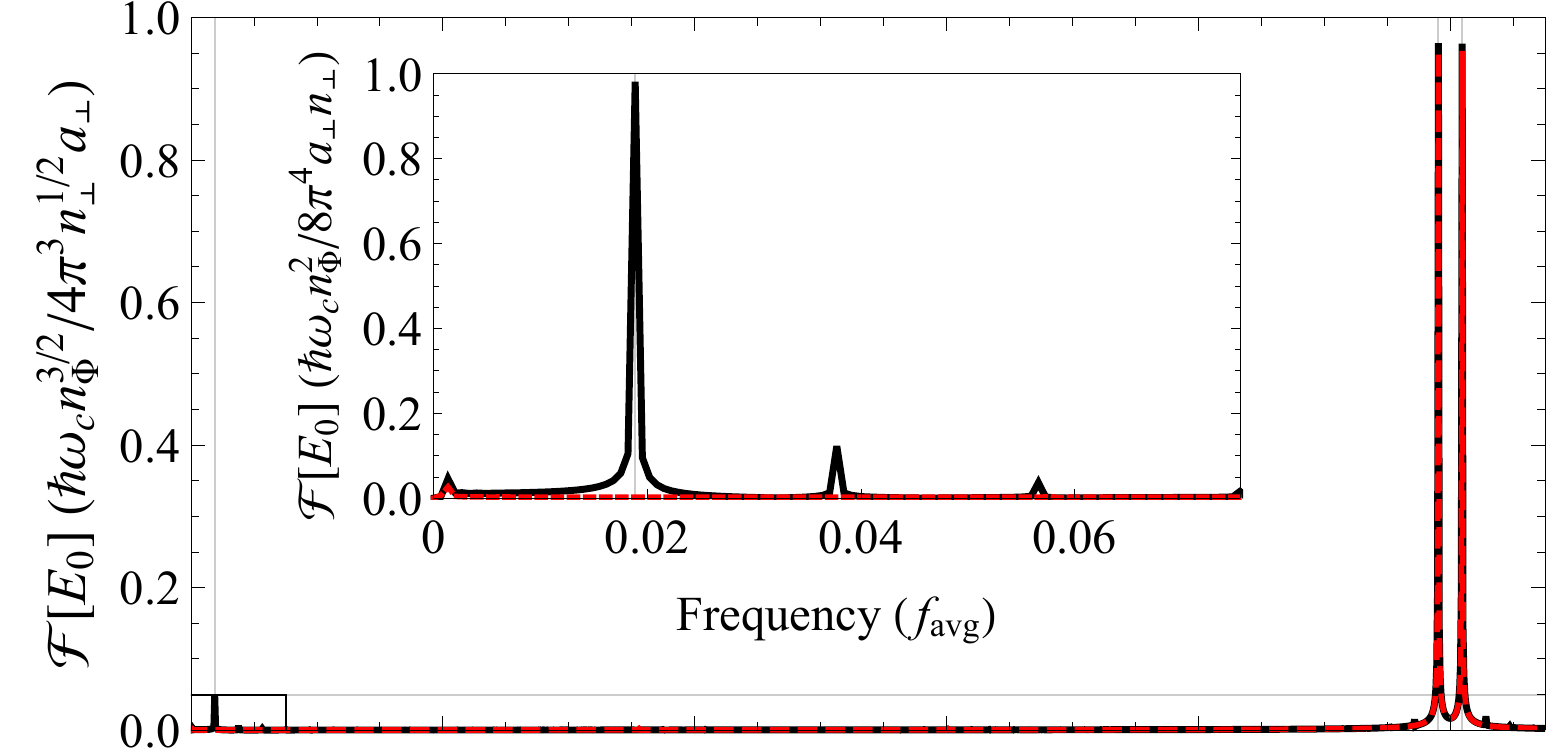}\\
    \includegraphics[width=\columnwidth]{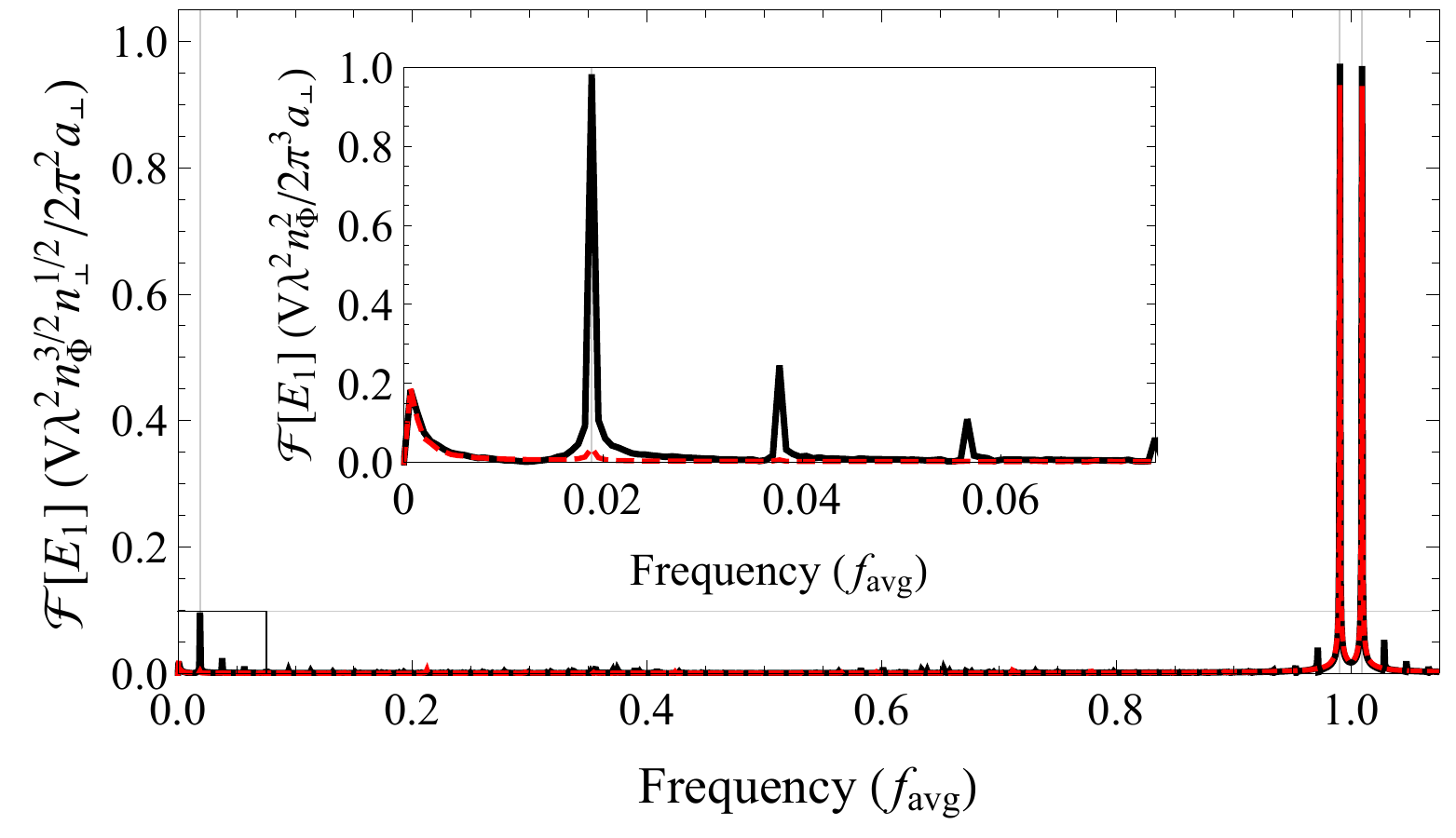}
    \caption{(Color online) The Fourier spectra for the 3D multi-layer system kinetic energy $F\text{3D}$ (upper panel) and Fock interaction energy $F_\text{3D}^{(1)}$ (lower panel).
    The red dashed lines show the same spectra for the case of fixed chemical potential~\cite{Supplement}.
    Frequencies are given in units of $f_\text{avg} = (f_++f_-)/2 = m^\ast\mu/e\hbar$. 
    We use $\delta f/f_\text{avg} = 4t_\perp/\mu = 0.0189$, consistent with the parameters of $\text{PtCoO}_2$~\cite{Arnold2020}.
    Vertical and horizontal guidelines mark the location and size of the largest difference frequency peak in each panel. %, and the horizontal lines mark the calculated size of the difference frequency peak.
    The insets of each panel expand the boxed sections of the main figures.
    %The small features at nearly 0 frequency are artifacts of imperfect subtraction of a polynomial background.
    The vertical scale of each panel is set by the appropriate amplitude in \cref{eq:F3D0osc} or \cref{eq:F3D1osc}.
    }
    \label{fig:3DFourier}
    \vspace{-15pt}
\end{figure}

We verify our analytic calculations by analyzing the clean, $T\to0$ limit of the system numerically.
The chemical potential to fix $n$ is evaluated numerically, the full $F_\text{3D}^{(0)}$ and $F_\text{3D}^{(1)}$ are evaluated on a set of points evenly spaced in $1/B$, a polynomial background is subtracted off, and the spectral content of the resulting oscillatory data is analysed via discrete Fourier transform.
The resulting Fourier spectra are presented in \cref{fig:3DFourier}.
The amplitudes at $f_\pm$ and $\delta f$ given by this analysis are found to closely match the $\tau_\text{qp}\to\infty$, $T\to0$ limit of the analytic results, Eqs.~\eqref{eq:F3D0osc} and \eqref{eq:F3D1osc}. 

\emph{Discussion} -- In the above analysis we have determined the oscillations of the free energy of the 3D system. The oscillatory part of the magnetization can be obtained as $\widetilde{M} = -\partial \widetilde{F}_\text{3D}/\partial B$, and naturally separates into interacting and non-interacting parts in the same way as $F_\text{3D}$, $\widetilde{M} = \widetilde{M}^{(0)} + \widetilde{M}^{(1)}$. 
The largest contributions arise from the derivative acting on the oscillating factors themselves, not the preceding amplitudes, as long as $k_B T, \hbar/\tau_\text{qp} \ll t_\perp$. 
Thus, the oscillation amplitudes of $\widetilde{M}$ can be acquired from those of $\widetilde{F}_\text{3D}$ in Eqs.~\eqref{eq:F3D0osc} and \eqref{eq:F3D1osc} by simply multiplying each term with a factor of $2\pi f/B^2$, where $f$ is the oscillation frequency of that term.

To compare the predictions of this theory with the experimental observations on $\text{PtCoO}_2$, we use the parameters reported in Ref.~\onlinecite{Arnold2020}~\footnote{Explicitly, we use $m^\ast = 1.18 m_e$, with $m_e$ the free electron mass, $\mu = \SI{2.963}{eV}$, $t_\perp = \SI{13.98}{meV}$, and c-axis lattice constant $a_\perp = \SI{17.808}{\angstrom}$.}.
Their investigation into short-range interactions in this material suggests an on-site Hubbard-like repulsion with $U\simeq \SI{6}{eV}$. 
%\NRCold{add ref.} \AAA{This is in the same Ref. \onlinecite{Arnold2020}. I changed wording a bit to make that clear.}
We take this to suggest short-range Coulomb interaction energies in general, including $V$, are on the scale of $\SI{}{eV}$, and use $\lambda^2$ on the scale of the in-plane area of the $\text{PtCoO}_2$ unit cell.
With these parameters, we find that the difference frequency amplitude of $\widetilde{M}^{(0)}$ is less than that of $\widetilde{M}^{(1)}$ for all fields $B < 4\pi V\lambda^2n_\perp m^\ast/\hbar e \sim \SI{600}{T}$.
The experimentally relevant low-field regime is far below this threshold, so within our theory dHvA oscillations at $\delta f$ are dominated by the interaction contribution, $\widetilde{M}^{(1)}$.
This amplitude is given by $\vert \widetilde{M}^{(1)} \vert_{\delta f} = V\lambda^2\, \delta f R_{D,1}^2 R_{T,1}^2/(\pi^2 \Phi_0^2 a_\perp)$. 
Note that this depends on magnetic field strength only through the squares of the Dingle and LK temperature factors, while the corresponding amplitude of the non-interacting term $\widetilde{M}^{(0)}$ depends on an additional factor of $B$ itself. 
This form of $\vert \widetilde{M}^{(1)} \vert_{\delta f}$ qualitatively matches the experimental results of the field-dependence of the low-frequency oscillations in Ref.~\onlinecite{Arnold2020}.
Furthermore, adding a factor of $2$ due to spin, the magnitude of this term is consistent with the measurements~\footnote{E. Hassinger (private communication)}.
Future measurements, focusing on the quantitative size of this low-frequency component, have the potential to allow a better understanding of the strength and nature of interlayer interactions.

The careful reader may note that the inequality $\delta f \ll f_\pm$, inherited from $t_\perp \ll \mu$, suggests that the dHvA oscillations our theory predicts at $\delta f$ should always be much smaller than those at $f_\pm$. 
This is in contrast with what is seen in experiment.
However, this apparent conflict may be reconciled by considering small-angle scattering off long-range disorder, e.g.\ spacial inhomogeneity of $\mu$.
It is well understood that long-range disorder suppresses QO amplitudes, and that the mean free path determined from Dingle measurements can be smaller than what is found in transport, which is less sensitive to small-angle scattering.
Indeed this is seen in experiment~\cite{Arnold2020}, where the mean free path determined from Dingle measurements is about ten times smaller than that found in transport.
This significant small-angle scattering strongly suppresses main frequency oscillations, however the difference frequency effect we find, being independent of $\mu$, is insensitive to this sort of disorder effect, allowing the two to be of a comparable scale. The starkly different dependence on small-angle scattering suggests that some measure of the strength of long-range disorder may be obtained by comparing Dingle measurements of difference and main frequency oscillations.

In summary,  by considering oscillations of the self-energy that are neglected in standard theories\cite{Luttinger1961,Shoenberg1984,Wasserman1996}, we have shown that Coulomb interactions can give rise to a new form of low-frequency dHvA oscillations in high mobility layered materials. These difference frequency oscillations are highly sensitive to the short-range disorder, their amplitude being suppressed by the square of the Dingle factor, but are insensitive to long-range disorder. They  can therefore become a dominant feature of oscillations in very high-mobility materials such as the delafossites.
Our theory shows that the size of the difference frequency oscillation is a measure of the strength and form of the interactions beyond the on-site Hubbard $U$. 
Although we have focused  on simple layered metals, our results suggest that dHvA studies may be of use to probe properties beyond just Fermi surface geometry in more general bilayer and multi-layer systems.

\begin{acknowledgments}
We thank Elena Hassinger and Andrew Mackenzie for helpful discussions,  for sharing experimental data, and for comments on an earlier draft.
This work is supported by EPSRC Grant No. EP/P034616/1 and by a Simons Investigator Award.
\end{acknowledgments}

\bibliography{references}

%%%%% Uncomment for arxiv submission
%%%%%% vvvvvvvvvvvvvvv

%%%%% Comment out for arxiv submission
%%%%%% VVVVVVVVVVVVVVV
% \documentclass[aps,prl,notitlepage,superscriptaddress]{revtex4-1}
% \usepackage{amsmath}
% \usepackage{amssymb}
% \usepackage{graphicx}% Include figure files
% \usepackage{bm}% bold math
% \usepackage{braket}
% \usepackage{hyperref}
% \usepackage[usenames]{xcolor}
% \usepackage[capitalize]{cleveref}
% \usepackage{siunitx}
% \usepackage{physics}

% \DeclareMathOperator{\sgn}{sgn}
%\setlength{\tabcolsep}{5pt}
%\renewcommand{\arraystretch}{2}

% \begin{document}
% \title{Supplement to Low-frequency Quantum Oscillations from Interactions in Layered Metals}

% \author{Andrew A. Allocca} \email{aa2182@cam.ac.uk}
% \author{Nigel R. Cooper}
% \affiliation{T.C.M. Group, Cavendish Laboratory, University of Cambridge, JJ Thomson Avenue, Cambridge, CB3 0HE, U.K.}

% \date{\today}
% \maketitle
%%%%%% ^^^^^^^^^^^^^^^^^^^^^^^^^^^^^^^
%%%%% Comment out for arxiv submission

%%%%% Uncomment for arxiv submission
%%%%%% VVVVVVVVVVVVVVV

\clearpage
\onecolumngrid
\appendix*

%%%%%% ^^^^^^^^^^^^^^^^^^^^^^^^^^^^^^^
%%%%% Uncomment for arxiv submission

\section{Grand Potential}
The method of calculating the grand potential of a non-interacting multi-layer system is presented in Ref~\onlinecite{Champel2001}.
Ignoring the effect of spin and assuming $\hbar\omega_c \ll 2t_\perp$ we have
\begin{multline}
    \Omega_\text{3D}^{(0)}(B,\mu) = -k_BT \Tr\log\left(-\beta\hat{G}^{-1}_0\right) = -k_B T n_\Phi \int_{k_z} \int d\epsilon \log\left(1+e^{-\epsilon/k_BT}\right) \sum_{l=0}^\infty \mathcal{A}(\epsilon-\xi_l(k_z))\\
    \approx -\frac{\nu_0}{2a_\perp}\left(\mu^2+2t_\perp^2\right) - \frac{\pi^2\nu_0}{6a_\perp}\left(k_BT\right)^2 + \frac{\nu_0}{a_\perp}\frac{(\hbar\omega_c)^2}{4\pi^3}\sqrt{\frac{\hbar\omega_c}{2t_\perp}}\sum_{p=1}^\infty\frac{(-1)^p}{p^{5/2}}R_{D,p}R_{T,p} \sum_{\alpha=\pm}\cos\left(2\pi p \frac{f_\alpha}{B} - \frac{\alpha\pi}{4}\right),
\end{multline}
where $\nu_0 = m^\ast/(2\pi\hbar^2)$ is the 2D density of states per spin at zero field.

The contribution to the grand potential at first order in the interlayer interaction can be simply calculated as well.
We use the part of the Hartree-Fock self-energy that is diagonal in Landau level index,
\begin{equation}
    \Sigma(k_z,B) = V\lambda^2 a_\perp n(B) - V\lambda^2 a_\perp \chi(B) \cos(k_za_\perp).
\end{equation}
Because we are calculating the contributions to $\Omega$ only up to first order in $V$ this approximation gives the correct result; directly computing the diagrams for the Hartree and Fock contributions to $\Omega$ gives the same result as we find below, though doing so obscures the point of departure from Ref.~\onlinecite{Luttinger1961}. 
The first order interaction term is then
\begin{multline}
    \Omega_\text{3D}^{(1)}(B,\mu) = k_B T \Tr\left(\hat{G}_0\hat{\Sigma}\right) + \frac{k_B T}{\mathbb{V}}\expval{S_I} \\
    = V\lambda^2 a_\perp k_B T n_\Phi \int_{k_z}\int d\epsilon\,n_F(\epsilon)\sum_{l=0}^\infty\mathcal{A}(\epsilon-\xi_l(k_z)) \left[n(B)-\chi(B) \cos(k_za_\perp)\right] + \frac{k_B T}{\mathbb{V}}\expval{S_I}\\
    = V\lambda^2 a_\perp \left[n(B)^2 - \chi(B)^2\right] + V\lambda^2 a_\perp \frac{n_I}{a_\perp}\left(\frac{n_I}{a_\perp} - 2n(B)\right) = V\lambda^2a_\perp\left[\left(n(B)-\frac{n_I}{a_\perp}\right)^2 - \chi(B)^2\right],
\end{multline}
where $\mathbb{V} = A L_z$ is the volume of the system.
In writing this result we have incorporated the contribution from the interaction with the ionic lattice, which contributes only at first order in $V$.
It is clear from this form that if the electron density $n$ is set equal to $n_I/a_\perp$ then the first term in the final expression exactly vanishes, with the Hartree term cancelling against contribution from the ionic lattice. 

\section{Fixing Electron Density}

When holding the particle density $n$ constant, the chemical potential necessarily becomes a function of the applied field, $\mu=\mu(B)$.
We can get some insight into the form of $\mu(B)$ by first examining how the density changes for fixed $\mu$ in a non-interacting single layer.  
The degeneracy of each Landau level is $N_\Phi = A B/\Phi_0$, so the number of electrons in each level increases with $B$. 
Therefore the total particle number for low temperature, $k_B T\ll\hbar\omega_c$, and fixed $\mu$ suddenly decreases by $N_\Phi$ every time a Landau level crosses above the Fermi energy, then rises again as the degeneracy of the still-filled states continues to increase.
The overall result is for $n(B)$ to oscillate around $n = n(B=0)$ in a sawtooth pattern.

To keep the density $n$ constant instead, the chemical potential must ``stick'' to a single Landau level as it moves in energy--the filling fraction of the highest level decreases to compensate for increasing degeneracy.
When the top level is completely emptied, $\mu(B)$ then jumps to the level next highest in energy.
We see that the chemical potential must then oscillate in a sawtooth pattern around its value at $B=0$, with the value of $\mu(B)$ equal to the energy of some particular Landau level.

\subsection{The Impact of Interactions on the Chemical Potential}

The above intuition ignores interactions, which in general will non-uniformly redistribute electrons between the levels of the system and therefore affect how the chemical potential must oscillate in order to keep the total density fixed.
Consider now an interacting system.
The grand potential may be written as
\begin{equation}
    \Omega(B,\mu) = \Omega^{(0)}(B,\mu) + \Omega_V(B,\mu),
\end{equation}
where $\Omega_V$ contains all interactions, so that $\eval{\Omega}_{V=0} = \Omega^{(0)}$.
The electron density in the system is
\begin{equation}
    n(B,\mu) = -\frac{\partial\Omega(B,\mu)}{\partial\mu} = -\frac{\partial\Omega^{(0)}(B,\mu)}{\partial\mu} -\frac{\partial\Omega_V(B,\mu)}{\partial\mu} \equiv n^{(0)}(B,\mu) + n_V(B,\mu).
\end{equation}

Define $\mu(B,n)$ to be the oscillatory chemical potential needed to exactly fix the electron density to the value $n$, i.e.\ $n(B,\mu(B,n)) \equiv n$.
We can separate $\mu(B,n)$ into oscillatory and non-oscillatory parts as
\begin{equation}
    \mu(B,n) = \mu_0(n) + \widetilde{\mu}(B,n) = \mu_0(n) + \widetilde{\mu}^{(0)}(B,n) + \widetilde{\mu}_V(B,n),
\end{equation}
where $\mu_0$ is the constant value of the chemical potential for $B=0$, and $\widetilde{\mu}$ is the full oscillatory component, with $\eval{\widetilde{\mu}}_{V=0} \equiv \widetilde{\mu}^{(0)}$, so that $\widetilde{\mu}_V$ contains all oscillations induced by the interaction.
Similarly we note $\eval{n^{(0)}(B,\mu_0 + \widetilde{\mu})}_{V=0} = n^{(0)}(B,\mu_0 + \widetilde{\mu}^{(0)})$ and define $n^{(0)}_V(B,\mu_0+\widetilde{\mu}) \equiv n^{(0)}(B,\mu_0 + \widetilde{\mu}) - n^{(0)}(B,\mu_0 + \widetilde{\mu}^{(0)})$ to be the entire part of $n^{(0)}(B,\mu_0+\widetilde{\mu})$ that depends on interactions.
The fixed-density condition then becomes
\begin{equation}
    n = n^{(0)}(B,\mu_0 + \widetilde{\mu}^{(0)}) + n^{(0)}_V(B,\mu_0 + \widetilde{\mu}) + n_V(B,\mu_0 + \widetilde{\mu}),
\end{equation}
with all effects of the interaction concentrated into the last two terms, marked with subscript $V$.
The value of $n$ is entirely independent of the interaction, equal to the ionic charge density in order for the system to be charge neutral overall.
For the fixed-density condition to be satisfied we must then have
\begin{gather}
    n = n^{(0)}(B,\mu_0 + \widetilde{\mu}^{(0)}) \label{eq:density}\\
    n^{(0)}_V(B,\mu_0 + \widetilde{\mu}^{(0)}) + n_V(B,\mu_0 + \widetilde{\mu}^{(0)} + \widetilde{\mu}_V) = 0.
\end{gather}

We now examine the free energy of this system, related to the grand potential by a Legendre transform.
Expanding $\Omega^{(0)}$ around the non-interacting parts of the chemical potential, we have
\begin{align}
    F(B,&n) = \Omega(B,\mu(B,n)) + n\,\mu(B,n) = \Omega^{(0)}(B,\mu(B,n)) + n\,\mu(B,n) + \Omega_V(B,\mu(B,n)) \nonumber\\
    &= \Omega^{(0)}(B,\mu_0 + \widetilde{\mu}^{(0)}) + \underbrace{\eval{\frac{\partial\Omega^{(0)}}{\partial\mu}}_{\mu_0+\widetilde{\mu}^{(0)}}}_{= -n} \widetilde{\mu}_V + \frac{1}{2} \eval{\frac{\partial^2\Omega^{(0)}}{\partial\mu^2}}_{\mu_0+\widetilde{\mu}^{(0)}} \widetilde{\mu}_V^2 + \dots + n\,\left(\mu_0 + \widetilde{\mu}^{(0)} + \widetilde{\mu}_V \right) + \Omega_V(B,\mu_0+\widetilde{\mu}) \nonumber\\
    &= \Omega^{(0)}(B,\mu_0 + \widetilde{\mu}^{(0)}) + n\,\left(\mu_0 + \widetilde{\mu}^{(0)}\right) + \Omega_V(B,\mu_0 + \widetilde{\mu}) + O(\widetilde{\mu}_V^2),
\end{align}
where in the second line we use \cref{eq:density} to relate the derivative of $\Omega^{(0)}$ evaluated at the non-interacting chemical potential to the fixed density $n$.
We see that the interaction appears only within $\Omega_V$ and in terms second order and higher in $\widetilde{\mu}_V$. 
Because $\Omega_V$ and $\widetilde{\mu}_V$ are themselves at least first order in the interaction, the only term in this expression that may occur at first order overall is obtained from the non-interacting part of the chemical potential inside $\Omega_V$. 
Therefore, if we discard contributions above first order, the free energy is
\begin{equation}
    F(B,n) \approx \Omega^{(0)}(B,\mu_0 + \widetilde{\mu}^{(0)}) + n\,\left(\mu_0 + \widetilde{\mu}^{(0)}\right) + \Omega^{(1)}(B,\mu_0 + \widetilde{\mu}^{(0)}),
\end{equation}
where $\Omega^{(1)}$ is the part of $\Omega_V$ first order in interactions.
Crucially, we see that $\widetilde{\mu}_V$ does not appear.
We conclude, then, that if we only consider the interaction up to first order it is unnecessary to consider how interactions affect the fixed-density condition, and the only relevant oscillatory part of the chemical potential is determined through \cref{eq:density}.

\subsection{Oscillatory Chemical Potential}

%An exact expression as above cannot be determined for finite quasiparticle lifetime or for nonzero temperature, nor does the above construction generalize straightforwardly to the 3D multi-layer system that is our primary interest.
%Instead, we
We follow Champel and Mineev~\cite{Champel2001} to derive an oscillatory chemical potential using the grand potential $\Omega_\text{3D}$.
As argued above, it suffices for our purposes to consider only the density coming from the non-interacting part of the grand potential.
This is
\begin{equation}
    n(B) = -\frac{\partial\Omega^{(0)}(B,\mu)}{\partial\mu} = \frac{\nu_0\mu}{a_\perp} + \frac{\nu_0}{a_\perp}\frac{\hbar\omega_c}{2\pi^2}\sqrt{\frac{\hbar\omega_c}{2t_\perp}}\sum_{p=1}^\infty \frac{(-1)^p}{p^{3/2}}R_{D,p}R_{T,p}\sum_{\alpha=\pm}\sin\left(2\pi p \frac{f_\alpha}{B}-\frac{\alpha\pi}{4}\right),
\end{equation}
which we demand to be equal to the density $n = n_I/a_\perp$ by letting $\mu = \mu(B,n)$.
Note that we arrive at exactly the same expression if we instead calculate $n(B)$ from the spectral density,  $n(B) = n_\Phi \int_{k_z}\int d\epsilon\,n_F(\epsilon)\sum_{l=0}^\infty\mathcal{A}(\epsilon-\xi_l(k_z))$. 
This then gives
\begin{equation}
    \mu(B,n) = \mu_0(n) + \frac{\hbar\omega_c}{2\pi^2} \sqrt{\frac{\hbar\omega_c}{2t_\perp}} \sum_{p=1}^\infty \frac{(-1)^{p+1}}{p^{3/2}} R_{D,p} R_{T,p} \sum_{\alpha=\pm} \sin\left(2\pi p \frac{\mu(B,n)+2\alpha t_\perp}{\hbar\omega_c} - \alpha\frac{\pi}{4}\right),
\end{equation}
where $\mu_0(n) = a_\perp n/\nu_0 = n_I /\nu_0$ in our system, and $R_{D,p}$ and $R_{T,p}$ are the Dingle factor and Lifshitz-Kosevich temperature factor defined in the main text.
In principle this equation determines $\mu(B,n)$ self-consistently.
However, we note that the oscillatory part is much smaller than the leading constant term because $\hbar\omega_c \ll 2t_\perp \ll \mu_0$ and $R_D,R_T \leq 1$, so better and better approximations to the exact form of $\mu(B,n)$ can be determined by iteration, starting with the replacement $\mu(B,n) \to \mu_0$ in the right-hand side of the equation.
For our purposes this starting point is a sufficient approximation, giving us
\begin{equation} \label{eq:mu3D}
    \mu(B,n) \approx \mu_0 + \frac{\hbar\omega_c}{2\pi^2} \sqrt{\frac{\hbar\omega_c}{2t_\perp}} \sum_{p=1}^\infty \frac{(-1)^{p+1}}{p^{3/2}} R_{D,p} R_{T,p} \sum_{\alpha=\pm} \sin\left(2\pi p \frac{f_\alpha}{B} - \alpha\frac{\pi}{4}\right) \equiv \mu_0 + \widetilde{\mu}(B).
\end{equation}
Indeed, if we were to consider the next iteration by inserting this form of $\mu(B,n)$ in place of $\mu_0$ within the sine, the additional terms we would acquire would be even smaller still and would only contribute very small corrections to our main results. 

\section{Free Energy Oscillations}

Free energy is related to the grand potential through a Legendre transformation,
\begin{equation}
    F_\text{3D}(B,n) = \Omega_\text{3D}(B,\mu(B,n)) + n\,\mu(B,n) = \underbrace{\Omega_\text{3D}^{(0)}(B,\mu(B,n)) + n\,\mu(B,n)}_{\equiv F_\text{3D}^{(0)}} + \underbrace{\Omega_\text{3D}^{(1)}(B,\mu(B,n))}_{\equiv F_\text{3D}^{(1)}}.
\end{equation}
Because $\mu(B,n)$ is independent of interactions, the free energy separates cleanly into non-interacting and interacting parts related to the corresponding parts of the grand potential, which we now analyze in turn focusing specifically on the oscillatory components at frequencies $f_\pm$ and $\delta f$.

We start with the non-interacting part.
We drop the terms of $\Omega^{(0)}$ that are independent of $\mu$ since they will not yield any oscillations after substituting $\mu\to\mu(B,n) = \mu_0 + \widetilde{\mu}(B)$.
We have
\begin{align}
    F_\text{3D}^{(0)} &= -\frac{\nu_0}{2a_\perp}\mu(B,n)^2 + \frac{\nu_0 \mu_0}{a_\perp}\mu(B,n) + \frac{\nu_0}{a_\perp}\frac{(\hbar\omega_c)^2}{4\pi^3}\sqrt{\frac{\hbar\omega_c}{2t_\perp}}\sum_{p=1}^\infty\frac{(-1)^p}{p^{5/2}}R_{D,p}R_{T,p} \sum_{\alpha=\pm}\cos\left(2\pi p \frac{\mu(B,n) + 2\alpha t_\perp}{\hbar\omega_c} - \frac{\alpha\pi}{4}\right) \nonumber \\
    &= \frac{\nu_0}{2a_\perp}\left(\mu_0^2 -\widetilde{\mu}^2\right) + \frac{\nu_0}{a_\perp}\frac{(\hbar\omega_c)^2}{4\pi^3}\sqrt{\frac{\hbar\omega_c}{2t_\perp}}\sum_{p=1}^\infty\frac{(-1)^p}{p^{5/2}}R_{D,p}R_{T,p} \sum_{\alpha=\pm}\left[\cos\left(2\pi p \frac{f_\alpha}{B} - \frac{\alpha\pi}{4}\right)\cos\left(2\pi p \frac{\widetilde{\mu}}{\hbar\omega_c}\right) \right. \nonumber\\
    & \hspace{10cm} \left.-\sin\left(2\pi p \frac{f_\alpha}{B} - \frac{\alpha\pi}{4}\right)\sin\left(2\pi p \frac{\widetilde{\mu}}{\hbar\omega_c}\right) \right].
\end{align}
The dominant contributions to oscillations at the frequencies of interest arise from the $p=1$ term of this sum since the size of the summand decreases quickly with growing $p$.
The same is true of the sum in the definition of $\widetilde{\mu}$ itself, allowing the same approximation. 
Because $\hbar\omega_c \ll 2t_\perp$ and $R_D,R_T \leq 1$ we can expand the factors involving sine and cosine of $\widetilde{\mu}$, keeping just the single largest term in each case, 
\begin{align}
    F_\text{3D}^{(0)} &\approx \frac{\nu_0}{2a_\perp}\left(\mu_0^2 -\widetilde{\mu}^2\right) - \frac{\nu_0}{a_\perp}\frac{(\hbar\omega_c)^2}{4\pi^3}\sqrt{\frac{\hbar\omega_c}{2t_\perp}}R_{D,1}R_{T,1} \sum_{\alpha=\pm}\left[\cos\left(2\pi \frac{f_\alpha}{B} - \frac{\alpha\pi}{4}\right) \right.\hspace{4cm} \nonumber\\
    & \hspace{5cm} \left. - \sin\left(2\pi \frac{f_\alpha}{B} -\frac{\alpha\pi}{4}\right) \frac{1}{\pi}\sqrt{\frac{\hbar\omega_c}{2t_\perp}} R_{D,1}R_{T,1} \sum_{\beta=\pm} \sin\left(2\pi\frac{f_\beta}{B}-\frac{\beta\pi}{4}\right) \right] \nonumber\\
    &= \frac{\nu_0}{2a_\perp}\mu_0^2 - \frac{\nu_0}{a_\perp}\frac{(\hbar\omega_c)^2}{4\pi^3}\sqrt{\frac{\hbar\omega_c}{2t_\perp}}R_{D,1}R_{T,1} \sum_{\alpha=\pm}\cos\left(2\pi \frac{f_\alpha}{B} - \frac{\alpha\pi}{4}\right) \nonumber\\
    &\hspace{2cm} + \frac{\nu_0}{a_\perp}\frac{(\hbar\omega_c)^2}{8\pi^4}\frac{\hbar\omega_c}{2t_\perp}R_{D,1}^2R_{T,1}^2 \sum_{\alpha,\beta=\pm}\sin\left(2\pi \frac{f_\alpha}{B} -\frac{\alpha\pi}{4}\right) \sin\left(2\pi\frac{f_\beta}{B}-\frac{\beta\pi}{4}\right). 
\end{align}
Trigonometric identities let us combine the product of sines in the last term, giving one term oscillating at the sum of $f_\alpha$ and $f_\beta$ and one term oscillating at their difference, which is either constant or equal to $\delta f$.
Discarding all constant terms and all terms with an oscillation frequency other than $f_\pm$ or $\delta f$, we are left with just $\widetilde{F}_\text{3D}^{(0)}$ as in \cref{eq:F3D0osc},
\begin{equation}
    \widetilde{F}_\text{3D}^{(0)}(B) = \frac{\hbar\omega_c n_\Phi^2}{8\pi^4 a_\perp n_\perp} R_{D,1}^2R_{T,1}^2 \sin\left(\frac{2\pi\,\delta f}{B}\right) - \frac{\hbar\omega_c n_\Phi}{4\pi^3a_\perp}\sqrt{\frac{n_\Phi}{n_\perp}}R_{D,1}R_{T,1}\sum_{\alpha=\pm}\cos\left(\frac{2\pi f_\alpha}{B}-\frac{\alpha\pi}{4}\right).
\end{equation}

Turning now to the interacting part of the free energy we have,
\begin{equation}
    F_\text{3D}^{(1)} = V \lambda^2 a_\perp\left[\left(n(B,\mu(B,n_I/a_\perp))-\frac{n_I}{a_\perp}\right)^2 - \chi(B,\mu(B,n_I/a_\perp))^2\right] = -V \lambda^2 a_\perp \chi(B,\mu(B,n_I/a_\perp))^2.
\end{equation}
Employing the same techniques as in the calculation of $\Omega$ we obtain
\begin{multline}
    \chi(B,\mu(B,n_I/a_\perp)) = n_\Phi \int_{k_z}\cos(k_z a_\perp)\int d\epsilon\,n_F(\epsilon) \sum_l \mathcal{A}(\epsilon-\xi_l(k_z)) \\
    = \frac{\nu_0 t_\perp}{a_\perp} + \frac{\nu_0}{a_\perp}\frac{\hbar\omega_c}{2\pi^2}\sqrt{\frac{\hbar\omega_c}{2t_\perp}} \sum_{p=1}^\infty \frac{(-1)^p}{p^{3/2}}R_{D,p}R_{T,p} \sum_{\alpha=\pm} \alpha \sin\left(2\pi p \frac{\mu(B,n_I/a_\perp) + 2\alpha t_\perp}{\hbar\omega_c}-\frac{\alpha\pi}{4}\right). 
\end{multline}
The constant term is much larger than the oscillatory term here, so for the free energy, in which this term appears squared, the dominant oscillatory contribution will result from the cross term. 
Keeping just this part, we have
\begin{align}
    F_\text{3D}^{(1)} &\approx -V\lambda^2 \frac{\nu_0^2 t_\perp}{a_\perp}\frac{\hbar\omega_c}{\pi^2}\sqrt{\frac{\hbar\omega_c}{2t_\perp}} \sum_{p=1}^\infty \frac{(-1)^p}{p^{3/2}}R_{D,p}R_{T,p} \sum_{\alpha=\pm} \alpha \sin\left(2\pi p \frac{\mu(B,n_I/a_\perp) + 2\alpha t_\perp}{\hbar\omega_c}-\frac{\alpha\pi}{4}\right) \nonumber\\
    & = -V\lambda^2 \frac{\nu_0^2 t_\perp}{a_\perp}\frac{\hbar\omega_c}{\pi^2}\sqrt{\frac{\hbar\omega_c}{2t_\perp}} \sum_{p=1}^\infty \frac{(-1)^p}{p^{3/2}}R_{D,p}R_{T,p} \sum_{\alpha=\pm} \alpha \left[\sin\left(2\pi p \frac{f_\alpha}{B}-\frac{\alpha\pi}{4}\right)\cos\left(2\pi p \frac{\widetilde{\mu}}{\hbar\omega_c}\right) \right. \nonumber\\
    & \hspace{9cm} \left.+ \cos\left(2\pi p \frac{f_\alpha}{B}-\frac{\alpha\pi}{4}\right)\sin\left(2\pi p \frac{\widetilde{\mu}}{\hbar\omega_c}\right)\right].
\end{align}
As above, the largest oscillatory contributions come from the $p=1$ term of this sum, and also for the sum within $\widetilde{\mu}$.
With these approximations and expanding the sine and cosine of $\widetilde{\mu}$ we have
\begin{multline}
    F_\text{3D}^{(1)} \approx V\lambda^2 \frac{\nu_0^2 t_\perp}{a_\perp}\frac{\hbar\omega_c}{\pi^2}\sqrt{\frac{\hbar\omega_c}{2t_\perp}} R_{D,1}R_{T,1} \sum_{\alpha=\pm} \alpha \sin\left(2\pi \frac{f_\alpha}{B}-\frac{\alpha\pi}{4}\right) \\
   + V\lambda^2 \frac{n_\Phi^2}{2\pi^3a_\perp} R_{D,1}^2R_{T,1}^2 \sum_{\alpha,\beta=\pm} \alpha \cos\left(2\pi p \frac{f_\alpha}{B}-\frac{\alpha\pi}{4}\right) \sin\left(2\pi\frac{f_\beta}{B}-\frac{\beta\pi}{4}\right) 
\end{multline}
The first term gives oscillations at $f_\pm$.
The trigonometric functions of the second term can be combined to give oscillations at the sum or difference of $f_\alpha$ and $f_\beta$.
Keeping just the terms oscillating at $f_\pm$ or $\delta f$ we then obtain $\widetilde{F}_\text{3D}^{(1)}$ as in \cref{eq:F3D1osc},
\begin{equation}
    \widetilde{F}_\text{3D}^{(1)}(B) = V\lambda^2 \frac{n_\phi^2}{2\pi^3a_\perp}R_{D,1}^2 R_{T,1}^2\cos(\frac{2\pi\,\delta f}{B}) + V\lambda^2 \frac{n_\Phi n_\perp}{2\pi^2a_\perp}\sqrt{\frac{n_\Phi}{n_\perp}}R_{D,1}R_{T,1}\sum_{\alpha=\pm} \alpha \sin\left(\frac{2\pi f_\alpha}{B}-\frac{\alpha\pi}{4}\right).
\end{equation}

\section{Comparison of Oscillations of F and $\Omega$}
As a point of comparison, we provide the oscillation amplitudes at the frequencies $f_\pm$ and $\delta f$ in both the case of fixed chemical potential $\Omega_\text{3D}$ and fixed electron density $F_\text{3D}$, further subdivided into interacting and non-interacting contributions.
The expressions are written in terms of the densities $n_\Phi = B/\Phi_0 = \nu_0 \hbar\omega_c$ and $n_\perp = 2\nu_0 t_\perp$, as well as the Dingle and LK temperature factors. 
We see that the the restriction to fixed $n$ does not affect the amplitude of oscillations at $f_\pm$, but it is responsible for the generation of large difference frequency oscillations.
For fixed chemical potential there are no oscillations at $\delta f$ at all in the non-interacting case, and in the interacting case they are very small, $n_\Phi/n_\perp = \hbar\omega_c/2t_\perp \ll 1$ smaller than in the case of fixed density. 

\begin{table}[h]
    \centering
    \vspace{5pt}
    \begin{tabular}{|c|c|c|}
    \hline
    & Amplitude at $\delta f$ & Amplitude at $f_\pm$ \\
    \hline
    \hline
    $\Omega_\text{3D}^{(0)}$ & 0 & $\displaystyle\frac{\hbar\omega_c n_\Phi}{4\pi^3 a_\perp} \sqrt{\frac{n_\Phi}{n_\perp}} R_{D,1} R_{T,1}$ 
    \\
    \hline
    $F_\text{3D}^{(0)}$ & $\displaystyle\frac{\hbar\omega_c n_\Phi^2}{8\pi^4a_\perp\,n_\perp} R_{D,1}^2 R_{T,1}^2$ & $\displaystyle\frac{\hbar\omega_c n_\Phi}{4\pi^3a_\perp}\sqrt{\frac{n_\Phi}{n_\perp}}R_{D,1} R_{T,1}$ 
    \\
    \hline
    \hline
    $\Omega_\text{3D}^{(1)}$ & $\displaystyle V\lambda^2\frac{n_\Phi^3}{4\pi^4a_\perp n_\perp} R_{D,1}^2 R_{T,1}^2$ & $\displaystyle V\lambda^2 \frac{n_\Phi n_\perp}{2\pi^2 a_\perp}\sqrt{\frac{n_\Phi}{n_\perp}}R_{D,1} R_{T,1}$  
    \\
    \hline
    $F_\text{3D}^{(1)}$ & $\displaystyle V\lambda^2 \frac{n_\Phi^2}{2\pi^3a_\perp} R_{D,1}^2 R_{T,1}^2$ & $\displaystyle V\lambda^2 \frac{n_\Phi n_\perp}{2\pi^2 a_\perp}\sqrt{\frac{n_\Phi}{n_\perp}}R_{D,1} R_{T,1}$ 
    \\
    \hline
    \end{tabular}
    \caption{The approximate amplitudes for the oscillations of the thermodynamic potentials $\Omega$ and $F$ of the 3D multi-layer system at frequencies $\delta f$ and $f_\pm$.}
    \label{tab:3Damps}
\end{table}

%\bibliography{supplement_references}
%%%%% Comment out for arxiv submission
%%%%%% VVVVVVVVVVVVVVV

% \bibliography{references}

% \end{document}

%%%%%% ^^^^^^^^^^^^^^^^^^^^^^^^^^^^^^^
%%%%% Comment out for arxiv submission

%%%%%% ^^^^^^^^^^^^^^^
%%%%% Uncomment for arxiv submission

\end{document}